\documentclass[12pt]{article}
\usepackage{latexsym}
\usepackage{amssymb, amsmath, xspace, lscape,  latexsym}

\newcommand{\G}{\textsf{G}}

\newcommand{\pr}{\prime}

\newcommand{\bea}{\begin{eqnarray}}
\newcommand{\eea}{\end{eqnarray}}
\def\beq#1#2\eeq{
        \begin{equation}
        \label{#1}
            #2
        \end{equation}}

\renewcommand{\H}{\textsf{H}}

\newcommand{\al}{\alpha}
\newcommand{\la}{\lambda}
\newcommand{\bt}{\beta}

\newcommand{\re}{\textrm{e}}

\renewcommand{\tilde}{\widetilde}

\def\btheor#1\etheor{
        \begin{theor}
            #1
        \end{theor}
    }

    \def\bsled#1\esled{
        \begin{sled}
            #1
        \end{sled}   }

\def\btheor#1\elemma{
        \begin{lemma}
            #1
        \end{lemma}
    }

    \def\bsled#1\esled{
        \begin{sled}
            #1
        \end{sled}   }

\newtheorem{theorem}{Theorem}

\newtheorem{lemma}{Lemma}

\def\hm#1{#1\nobreak\discretionary{}{\hbox{\m@th$#1$}}{}}
\def\mi#1{\discretionary{\hbox{\m@th$#1$}}{\hbox{\m@th$#1$}}{}}

\vspace{4ex}
\begin{document}
\title{\bf Painlev\'e V and the distribution function of a discontinuous
linear statistics in the Laguerre Unitary Ensembles}
\author{Estelle Basor\thanks{ebasor@aimath.org.
          Supported in part by NSF Grant DMS-0500892}\\
        American Institute of Mathematics\\
        Palo Alto, California\\
        94306 USA \\
        \and
        Yang Chen\thanks{ychen.ic.ac.uk}\\
        Department of Mathematics\\
        Imperial College London\\
        180 Queen's Gates\\
        London SW7 2BZ UK}
\date{28-10-2008}
\maketitle
\begin{abstract}
In this paper we study the characteristic or generating function of a certain
discontinuous linear statistics of the Laguerre unitary ensembles
and show that this is a particular fifth Painl\'eve transcendant
in the variable $t,$ the position of the discontinuity.

The proof of the ladder operators adapted to orthogonal polynomial
with discontinuous weight announced sometime ago \cite{chen3} is presented here,
followed by the non-linear difference equations satisfied by
two auxiliary quantities and the derivation of the Painl\'eve equation.

\end{abstract}

\vfill\eject

\setcounter{equation}{0}
\section{Introduction}

In the theory of random matrix ensembles with unitary symmetry the
real eigenvalues $\{x_j\}_{j=1}^{n}$ have the joint probability
distribution
\bea \textsf{P}(x_1,...,x_n)dx_1...
dx_n=\frac{1}{n!\:D_n} \prod_{1\leq j<k\leq
n}(x_k-x_j)^2\prod_{l=1}^{n}w_0(x_l)dx_l,
\eea
where $w_0(x)$ with
$x\in[a,b]$ say, is strictly positive and satisfies a Lipshitz
condition and has finite moments, that is, the existence of the integrals,
$$
\int_{a}^{b}x^jw_0(x)dx,\;\;j\in\{0,1,2,...\}.
$$
Here $D_n$ is the normalization constant \bea
D_n[w_0]=\frac{1}{n!}\int_{[a,b]^n}\prod_{1\leq j<k\leq
n}(x_k-x_j)^2\prod_{l=1}^{n}w_0(x_l)dx_l, \eea so that \bea
\int_{[a,b]^n}\textsf{P}(x_1,...,x_n)dx_1...dx_n=1.
\eea
We include the cases where  $a$ may be $-\infty$ and/or $b$ may be $\infty$. For a comprehensive
study of the theory of random matrices see \cite{MM}.

A linear statistics is a linear sum of a certain function $g$ of the
random variable $x_j:$
\bea \sum_{k=1}^{n}g(x_k).\eea
The generating function of such a linear statistics is,
by definition the average of $\exp(\la\sum_kg(x_k)),$ with respect to
the joint probability distribution (1.1), where $\la$ is a parameter, reads,
\bea
\int_{[a,b]^n}\exp\left[\la\sum_{k=1}^{n}\:g(x_k)\right]
\textsf{P}(x_1,...,x_n)\prod_{k=1}^{n}dx_k.
\eea
More generally, we can consider
\bea \int_{[a,b]^n}\left[\prod_{k=1}^{n}
f(x_{k})\right] \textsf{P}(x_1,...,x_n)\prod_{k=1}^{n}dx_k
\eea

If
$f$ is a "smooth" function then asymptotic formulas for large $n$
for the characteristic functions have been obtained for the Hermite
case, where $w_0(x)=\re^{-x^2},\;x\in(-\infty,\infty)$ by Kac \cite{Kac} and
Akhiezer\cite{Akh} and generalized by many authors. See \cite{BS2} for a
history of this problem. These results are continuous analogs of the
classical Szeg\"o limit theorem on Toeplitz determinants

 In the  Laguerre case where $w_0(x)=x^{\al}\re^{-x},\:x\in[0,\infty)$ an
analogous formula was found recently for smooth functions in \cite{BCW}. However, results for the
situations where
$f$ has discontinuities are harder to come by. We mention here the original studies in \cite{BW}
where $f$ has several discontinuities and which corresponds to the Hermite case. More general results
can be found in \cite{BSW, Arch, BW2} and \cite{Mik}. Also, results that correspond to $\alpha = \pm 1/2$
and large $n$ appear in \cite{BEW}.

In this paper we investigate the case where $n$ is finite, and $f$ is
constant except for a jump at $t \in[0,\infty),$
and is of the form
\bea
f(x,t) = A + B \theta( x-t)
\eea
 where $\theta(x)$ is one for $x > 0$ and zero otherwise and $A \geq 0$ and $B >0$.
 In the special case of linear statistics the function $g$ will take the form
\bea
g(x,t):=\theta(x-t)\:\ln\left(1+\frac{\bt}{2}\right)+\theta(t-x)\:\ln\left(1-\frac{\bt}{2}\right),
\eea
where $-1<\frac{\bt}{2}<1.$
This corresponds to a function $f$
where
\[f(x,t) = \left(1-\frac{\bt}{2}\right)^{\la}+\left[\left(1+\frac{\bt}{2}\right)^{\la}-
\left(1-\frac{\bt}{2}\right)^{\la}\right]\theta(x-t) \]
that is
\[ A = \left(1-\frac{\bt}{2}\right)^{\la} \]
and
\[ B = \left(1+\frac{\bt}{2}\right)^{\la}-
\left(1-\frac{\bt}{2}\right)^{\la}. \]
We also point out that if $A = 0$ and $B = 1$  then we have the important case where we are
computing the probability that all the eigenvalues are in the interval $[t, \infty).$ This case
of course is
what leads to the now well-known Tracy-Widom laws. More will be said about this later.

Our main tool will be to use the theory of orthogonal polynomials. Previously,  in random matrix theory
one made use of the orthogonal polynomials associated to the weight that defined the ensemble.
Fundamental quantities were then described in terms of Fredholm determinants. While both the authors
are very fond of determinants, in this work, we do not consider Fredholm determinants. Instead we
consider the polynomials that are orthogonal to the perturbed weight, that is a regular or "nice" weight
multiplied by the discontinuous factor given in (1.7). In this manner we are able to use
the results of the  orthogonal polynomials to derive equations associated with the various statistics of
interest.

The idea is that we write the multiple integral in (1.6) as a Hankel determinant. We then need to know
information about the norms of the orthogonal polynomials. To understand this we need to know
something about the recursion coefficients of the polynomials. This will lead us naturally to another
pair of auxiliary
quantities that depend on $t$ and $n$. In the paper they are called $r_{n}$ and $R_{n}$.
Using these auxiliary quantities we are able to produce the second order non-linear
differential equations satisfied by $S_n=1-1/R_n$ which
turns out to be a particular fifth Painl\'eve transcendent, in addition to the Jimo-Miwa-Okamoto
$\sigma$ form \cite{Jimbo} satisfied by the logarithmic derivative of the Hankel determinant with
respect to $t.$ We also wish to emphasize that the logarithmic derivative of the Hankel determinant can be computed very naturally in terms of our quantity $r_{n}(t)$ and its derivative and the relationships between these quantities arise naturally using this approach.

We also derive a discrete version of the
$\sigma$ form of a nonlinear second order difference equation satisfied by the same
logarithmic derivative. Our computations show that in fact
the values of our generalized polynomials at the end points of the intervals are intimately related to the
resolvent kernels found in the standard approach of Tracy and Widom. This is really not surprising, since
we are all starting with the same multiple integral. Rather, our point is that computations can all be made
by using only the very basic theory of orthogonal polynomials.

The Palinlev\'{e} equation can be found in \cite{TW}. The second order difference equation
[(4.29), {\bf Theorem 8}], as far as we know is a new equation.

 In the next section the proof for a pair of ladder operators,
and the associated supplementary conditions adapted to orthogonal polynomials with
discontinuous weights which was announced sometime ago \cite{chen3} will be provided. In section {\bf 3},
a system of difference equations satisfied by two auxiliary quantities $r_n$ and $R_n$ (these will ultimately
determine the recurrence coefficients for the orthogonal polynomials) are derived. In section {\bf 4}
we derive a second order non-linear differential equation which turns out to be a particular
fifth Painlev\'e transcendent. In the process we identity that the quantity
$$
S_n(t):=1-\frac{1}{R_n(t)},
$$
to be such an equation. Furthermore we show that the logarithmic derivative of the generating function
$$
H_n(t):=t\frac{d}{dt}\ln \textsf{G}(n,t)=t\frac{d}{dt}\ln D_n(t)
$$
satisfies both the continuous and discrete $\sigma$ form of Painlev\'e V.

\setcounter{equation}{0}
\section{Ladder operators and supplementary conditions}

 According the general theory of orthogonal polynomials of one variable, for a generic weight $w$,
the normalization constant (1.2) has the two more alternative representations
\bea
D_n[w]&:=&\det(\mu_{i+j})_{i,j=0}^{n-1}
:=\det\left(\int_{a}^{b}x^{i+j}\:w(x)dx\right)_{i,j=0}^{n-1}\\
&=&\prod_{j=0}^{n-1}\:h_j,
\eea
where the determinant of the moment matrix $(\mu_{i+j})$ is the Hankel determinant.
Here $\{h_j\}_{j=0}^{n}$ is the square of the $L^2$ norm
of the sequence of (monic-)polynomials $\{P_j(x)\}_{j=0}^{n}$ orthogonal with respect to
$w$ over $[a,b];$
\bea
\int_{a}^{b}P_{i}(x)P_{j}(x)w(x)dx=\delta_{i,j}h_j.
\eea
Therefore with reference to (1.2) and (1.5) the quantity that we need to compute is
\[
\G(t,n) = \frac{D_n\left[ w \right]}{D_n[w_0]} = \frac{\prod_{i=0}^{n-1}h_i(t)}{\prod_{i=0}^{n-1}h_i},
\]
where $w(x,t):=x^{\al}\re^{-x}(A+B\theta(x-t))$ and $h_k(t)$ is defined by
\bea
\int_{0}^{\infty}\{P_k(x)\}^2
(A+B\:\theta(x-t) )
\:x^{\al}\:\re^{-x}dx=h_k(t).
\eea
We also denote
$$
D_n(t):= D_n[w(.,t)].
$$
This leads to the generic problem of the characterization of polynomials orthogonal with respect to
"smooth" weights $w_0(x)$ perturbed by a jump factor where the discontinuity is at $t.$ So if we write
\bea
w_J(x,t):=A+B\:\theta(x-t),\quad A\geq 0,\quad A+B>0
\eea
then
\bea
\int_{a}^{b}P_i(x)P_j(x)w_0(x)w_J(x,t)dx=\delta_{i,j}h_j(t).
\eea
It follows from the orthogonality relations that,
\bea
zP_n(z)=P_{n+1}(z)+\al_n(t)P_{n}(z)+\bt_n(t)P_{n-1}(z).
\eea
This three term recurrence relations, together
with the "initial" conditions,
$$
P_0(z)=1,\;\;\;\bt_0P_{-1}(z)=0,
$$
generates the monic polynomials,
\bea
P_n(z)=z^{n}+\textsf{p}_1(n,t)z^{n-1}+...
\eea
the first two of which are
\bea
P_0(z)&=&1\nonumber\\
P_1(z)&=&z-\al_0(t)=z-\frac{\mu_1(t)}{\mu_0(t)}.
\eea
Note that due to the $t$ dependence
of the weight, the coefficients of the polynomials and the
recurrence coefficients $\al_n$ and $\bt_n$ also depend on $t$ the position of the jump.
However, unless it is required we do not display the $t$ dependence.

From (2.7) and (2.8), we find, for $n\in\{0,1,2,..\}$
\bea
\al_n&=&\textsf{p}_1(n,t)-\textsf{p}_1(n+1,t),\nonumber\\
\sum_{j=0}^{n-1}\al_j&=&-\textsf{p}_1(n,t)
\eea
where $\textsf{p}_1(0,t):=0.$

From (2.6) and (2.7) we have the well-known strictly positive expression,
\bea
\bt_n:=\frac{h_n}{h_{n-1}}.
\eea
Another consequence of the recurrence relation is the Christoffel-Darboux formula
$$
\sum_{k=0}^{n-1}\frac{P_k(x)P_k(y)}{h_k}=\frac{P_{n}(x)P_{n-1}(y)-P_{n}(y)P_{n-1}(x)}{h_{n-1}(x-y)}.
\eqno(C-D)
$$
The above basic information about orthogonal polynomials can be found in \cite{Sze}.

In this section, we give an account of a recursive algorithm for the
determination of the $\al_n,\;\bt_n$ for a given weight. This is based on a pair of
ladder operators and the associated supplementary conditions to be denoted as
$(S_1)$ and $(S_2)$. For an general "smooth" weight the
lowering and raising operators has been derived by many authors \cite{Bonan,chen1,chen2, Shohat}.
We should like to note here A.P. Magnus's contribution to this formalism
\cite{Mag1, Mag2, Mag3}. Indeed, we have been motivated by the investigation of \cite{Mag3}
where he obtained the large $n$ behavior of the recurrence coefficients of a generalization
of the Jacobi polynomials in which the standard Jacobi weight is perturbed by a "line" analogue to
the Fisher-Hartwig singularity. We end the discussion about the ladder operators with the remark
that the supplementary conditions for orthogonal polynomials on the unit circle was found in \cite{Basor}
and have been used to compute explicitly the Toeplitz determinants with Fisher-Hartwig symbols.

The lemma below gives a detailed proof of the ladder operators in the
discontinuous case where the results were announced sometime ago \cite{chen3}.
\begin{lemma}
Let $w_0(x),\;\;x\in[a,b]$ be a smooth weight function where the associated moments,
\bea
\int_{a}^{b}x^{j}w_0(x)dx,\;\;j\in\{0,1,2,..\}
\eea
of all order exist.

Let $w_0(a)=w_0(b)=0,$ and $\textsf{v}_0(x):=-\ln w_0(x).$

The lowering and raising operators for
polynomials orthogonal with respect to
$$w(x):=w_0(x)w_J(x,t),$$
 are
\bea
P_n^{\pr}(z)&=&-B_n(z)P_n(z)+\bt_nA_n(z)P_{n-1}(z),\\
P_{n-1}^{\pr}(z)&=&[B_n(z)+\textsf{v}_0^{\pr}(z)]P_{n-1}(z)-A_{n-1}(z)P_n(z),
\eea
where
\bea
A_n(z)&:=&\frac{R_n(t)}{z-t}+\frac{1}{h_n}\int_{a}^{b}
\frac{\textsf{v}_0'(z)-\textsf{v}_0'(y)}{z-y}
P_n^2(y)w(y)dy\\
B_n(z)&:=&\frac{r_n(t)}{z-t}+
\frac{1}{h_{n-1}}\int_{a}^{b}\frac{\textsf{v}_0'(z)-\textsf{v}_0'(y)}{z-y}
P_n(y)P_{n-1}(y)w(y)dy\\
R_n(t)&:=&B\:\frac{w_0(t)}{h_n(t)}\{P_n(t,t)\}^2\\
r_n(t)&:=&B\:\frac{w_0(t)}{h_{n-1}(t)}P_n(t,t)P_{n-1}(t,t).
\eea
where
$$P_n(t,t):=P_n(z,t)\big|_{z=t}.$$

 Here $\ln w_0(x),$ is well defined since $w_0(x)$
is suppose to be strictly positive for $x\in[a,b].$
\end{lemma}
{\em Proof:} We start from
\bea
P_n'(z)=\sum_{k=0}^{n-1}C_{nk}P_k(z),\nonumber
\eea
where $C_{nk}$ is determined from the orthogonality relations,
\bea
C_{nk}=\frac{1}{h_k}\int_{a}^{b}P_n'(y)P_k(y)w(y)dy.\nonumber
\eea
Therefore
\bea
P_n'(z)&=&\sum_{k=0}^{n-1}\frac{P_k(z)}{h_k}\int_{a}^{b}P_n'(y)P_k(y)\:w(y)dy\nonumber\\
&=&-\sum_{k=0}^{n-1}\int_{a}^{b}\frac{P_k(z)}{h_k}P_n(y)
\{P_k'(y)w(y)+P_k(y)\left[B\delta(y-t)w_0(y)+w_0'(y)w_J(y,t)\right]\}dy\nonumber\\
&=&-\int_{a}^{b}P_n(y)\sum_{k=0}^{n-1}\frac{P_k(z)P_k(y)}{h_k}\left[B\:w_0(y)\delta(y-t)
+\frac{w_0'(y)}{w_0(y)}w(y)\right]dy
\nonumber\\
&=&-\int_{a}^{b}P_n(y)\sum_{k=0}^{n-1}\frac{P_k(z)P_k(y)}{h_k}\{B\:w_0(y)\delta(y-t)
+[\textsf{v}_0'(z)-\textsf{v}_0'(y)]w(y)\}dy\nonumber\\
&=&-\int_{a}^{b}P_n(y)\frac{P_{n}(z)P_{n-1}(y)-P_n(y)P_{n-1}(z)}{h_{n-1}(z-y)}
\{B\delta(y-t)w_0(y)+[\textsf{v}_0'(z)-\textsf{v}_0'(y)]w(y)\}dy\nonumber
\eea
where we have used integration by parts, (C-D), the definition of $\textsf{v}_0,$ (2.11)
and that
$$
\int_{a}^{b}P_n(y)P_{k}(y)w(y)dy=0,\quad k=0,1,2,...,n-1,
$$
to arrive at the above. A little simplification produces (2.15) and (2.16) follows
from straight forward application of the recurrence relations.$\quad\quad\Box$

\noindent
{\bf Remark 1.} If $w_0(a)\neq 0,$ $w_0(b)\neq 0,$ the terms
$$
w(y)\frac{\{P_{n}(y,t)\}^2}{h_n(t)(z-y)}\Bigg|_{y=a}^{b}\,\,\,\,   \mbox{and} \,\,\,\,\,\,
w(y)\frac{P_n(y,t)P_{n-1}(y,t)}{h_{n-1}(t)(z-y)}\Bigg|_{y=a}^{b}
$$
are to be added into the definition of $A_n(z)$ and $B_n(z)$ respectively.

\noindent
{\bf Remark 2.} If there are several jumps at $t_1,...,t_N$ then the first term of (2.15) and (2.16)
should be replaced by
\bea
&&\sum_{j=1}^{N}\frac{R_{n,j}(t_j;t)}{z-t_j}\nonumber\\
&&\sum_{j=1}^{N}\frac{r_{n,j}(t_j;t)}{z-t_j}\nonumber
\eea
where
\bea
R_{n,j}(t_j;t)&:=&B_j\frac{w_0(t_j)}{h_{n}(t)}\{P_n(t_j;t)\}^2\nonumber\\
r_{n,j}(t_j;t)&:=&B_j\frac{w_0(t_j)}{h_{n-1}(t)}P_n(t_j;t)P_{n-1}(t_j;t)\nonumber\\
t&:=&(t_1,...,t_N).\nonumber
\eea

As in the case of the smooth weight the "coefficients" $A_n(z)$ and $B_n(z)$ that appear in the ladder
operators satisfy two identities valid for all
$z\in\mathbb{C}\cup\{\infty\},$ which we gather in the next lemma.
\begin{lemma}
The functions $A_n(z)$ and $B_n(z)$ satisfy the following identities which hold for all $z:$
$$
B_{n+1}(z)+B_n(z)=(z-\al_n)A_n(z)-\textsf{v}_0'(z)\eqno(S_1)
$$
$$
1+(z-\al_n)[B_{n+1}(z)-B_n(z)]=\bt_{n+1}A_{n+1}(z)-\bt_{n}A_{n-1}(z)\eqno(S_2)
$$
\end{lemma}

{\em Proof:} By a direct computation using the definition of $A_n(z)$ and $B_n(z)$. $\quad\quad\Box$
\\
\\
It turns out that a suitable combination of $(S_1)$ and $(S_2)$ produces an identity
involving $\sum_{j=0}^{n-1}A_j(z),$ from which further insight into the recurrence
coefficients may be gained.

\begin{lemma}
$A_n(z),\;B_n(z)$ and $\sum_{j=0}^{n-1}A_j(z)$ satisfy the identity
$$
[B_n(z)]^2+\textsf{v}'_0(z)\;B_n(z)+\sum_{j=0}^{n-1}A_j(z)=\bt_nA_{n}(z)A_{n-1}(z)\eqno(S_2')
$$
\end{lemma}

{\em Proof:} Multiply $(S_2)$ by $A_n(z)$ and replace $(z-\al_n)A_n(z)$ in the resulting
equation by

 $B_{n+1}(z)+B_n(z)+\textsf{v}_0'(z).$ See $(S_1)$. This leads to
\bea
[B_{n+1}(z)]^2-[B_n(z)]^2+\textsf{v}_0'(z)[B_{n+1}(z)-B_n(z)]+A_n(z)=
\bt_{n+1}A_{n+1}(z)A_n(z)-\bt_nA_{n}(z)A_{n-1}(z).\nonumber
\eea
Taking a telescopic sum of the above equation from $0$ to $n-1$ with the "initial" conditions,
$B_0(z)=0$ and $\bt_0A_{-1}(z)=0,$ we have $(S_2').$ $\quad\quad \Box$

 Let $y=P_n(z)$ we find by eliminating $P_{n-1}(z)$ from the raising and lowering operators,
the second order differential equation

\begin{lemma}
\bea
y''(z)-\left(\textsf{v}_0'(z)+\frac{A_n'(z)}{A_n(z)}\right)y'(z)
+\left(B_n'(z)-B_n(z)\frac{A_n'(z)}{A_n(z)}+\sum_{j=0}^{n-1}A_j(z)\right)y(z)=0.
\eea
\end{lemma}

{\em Proof:} By a straight forward computation using (2.15), (2.16) and $(S_2')$. $\quad\quad\Box$

Recalling (2.17) and (2.18) we note that if $\textsf{v}_0'(z)$ is rational in $z$ then
the difference kernel, $[\textsf{v}_0'(z)-\textsf{v}_0'(y)]/(z-y)$ is rational
in $z$ and $y.$ Consequently $(S_1)$ and $(S_2')$ may be put to good use to obtain
a system of difference equations satisfied by the auxiliary quantities $R_n$ and $r_n$ and
the recurrence coefficients $\al_n $ and $\bt_n.$ This will be clear in the next section.

\setcounter{equation}{0}
\section{Recurrence coefficients and difference equations.}

For the problem at hand,
$$
w_0(x)=x^{\al}\re^{-x},\;\;x\in[0,\infty),
$$
$$
\textsf{v}_0(x):=-\ln w_0(x)=-\al\ln x+x
$$
and for $\al>0,$ $w_0(0)=0.$ Note that $w(\infty)=0.$  An easy computation gives,
\bea
\frac{\textsf{v}_0'(z)-\textsf{v}_0'(y)}{z-y}=\frac{\al}{zy}.\nonumber
\eea
Using these and integration by parts we have the following
\begin{lemma}
\bea
A_n(z)&=&\frac{R_n(t)}{z-t}+\frac{1-R_n(t)}{z},\\
B_n(z)&=&\frac{r_n(t)}{z-t}-\frac{n+r_n(t)}{z},\\
{\rm where\;}\nonumber\\
R_n(t)&:=&B\:w_0(t)\frac{\{P_{n}(t,t)\}^2}{h_n(t)}\\
r_{n}(t)&:=&B\:w_0(t)\frac{P_n(t,t)P_{n-1}(t,t)}{h_{n-1}(t)}.
\eea
\end{lemma}

{\em Proof:} Through integration by parts we find,
\bea
\al\int_{0}^{\infty}y^{\al-1}\re^{-y}w_J(y;t)\{P_n(y,t)\}^2dy&=&h_n(t)-B\:w_0(t)\{P_n(t,t)\}^2\\
\al\int_{0}^{\infty}y^{\al-1}\re^{-y}w_J(y;t)P_{n}(y,t)P_{n-1}(y,t)dy&=&
-n\:h_{n-1}(t)\nonumber\\
&-&B\:w_0(t)P_{n}(t,t)P_{n-1}(t,t),
\eea
and we have used the fact that
$$
\frac{\partial}{\partial x}P_n(x,t)=n\:P_{n-1}(x,t)+{\rm lower\;degree\;}
$$
to arrived at (3.6). From (3.5) and (3.6) and the definitions of $A_n(z)$ and $B_n(z),$
(3.1)---(3.4) follows.$\quad\quad\Box$

Substituting (3.1) and (3.2) into $(S_1)$ we find by equating the residues
\bea
r_{n+1}+r_n&=&R_n(t-\al_n)\\
-(r_{n+1}+r_n)&=&2n+1+\al-\al_n(1-R_n).
\eea
\begin{lemma}
\bea
\al_n&=&2n+1+\al+tR_n\\
r_{n+1}+r_n&=&R_n(t-\al_n)
\eea
\end{lemma}

{\em Proof:} (3.7)+(3.8) implies (3.9) and we restate (3.7) as (3.10).$\quad\quad\Box$

Substituting (3.1) and (3.2) into $(S_2'),$ we find, after some elementary but
messy computations,
\bea
[B_n(z)]^2+\textsf{v}_0'(z)\:B_n(z)&+&\sum_{j=0}^{n-1}A_j(z)=
\frac{r_n^2}{(z-t)^2}+\frac{(n+r_n)(\al+n+r_n)}{z^2}\nonumber\\
&+&\frac{\sum_{j=0}^{n-1}R_j+{r_n\left[1-\frac{\al}{t}
-\frac{2(n+r_n)}{t}\right]}}{z-t}\nonumber\\
&+&\frac{1}{z}\:\left[n-\sum_{j=0}^{n-1}R_j+(n+r_n)\left(\frac{2r_n}{t}-1\right)+\frac{\al r_n}{t}\right]
\eea
and
\bea
\bt_{n}A_{n}(z)A_{n-1}(z)&=&\frac{\bt_nR_nR_{n-1}}{(z-t)^2}+\frac{\bt_n(1-R_n)(1-R_{n-1}}{z^2}
\nonumber\\
&+&\frac{1}{t}\left(\frac{1}{z-t}-\frac{1}{z}\right)\:\bt_n
\left[(1-R_{n})R_{n-1}+(1-R_{n-1})R_n\right].
\eea
Now $(S_2')$ implies

\begin{lemma}
For a fixed $t,$ the quantities $r_n,\:R_n,\;\bt_n$ satisfy the equations
\bea
r_n^2&=&\bt_nR_nR_{n-1}\\
(n+r_n)(n+\al+r_n)&=&\bt_n(1-R_n)(1-R_{n-1})\\
\sum_{j=0}^{n-1}R_j+r_n\left[1-\frac{\al}{t}-\frac{2(n+r_n)}{t}\right]
&=&\frac{\bt_n}{t}\left[(1-R_n)R_{n-1}+(1-R_{n-1})R_n\right].
\eea
\end{lemma}

{\em Proof:} The equations (3.13)---(3.15) are obtained by equating residues of
$(S_2').$ $\quad\quad\Box$

In the next Lemma an expression is found for $\bt_n$ in terms of $r_n$ and $R_n.$

\begin{lemma} In terms of $r_n$ and $R_n,$ $\bt_n$ the
 off-diagonal recurrence coefficient reads
 \bea
 \bt_n=\frac{1}{1-R_n}\left[r_n(2n+\al)+n(n+\al)+\frac{r_n^2}{R_n}\right].
 \eea
\end{lemma}
{\em Proof:}
\noindent
We eliminate $\bt_nR_nR_{n-1}$ from (3.13) and (3.14) to find,
\bea
r_n(2n+\al)+n(n+\al)&=&\bt_{n}(1-R_n-R_{n-1})\\
&=&\bt_n(1-R_n)-\frac{r_n^2}{R_n}.
\eea
In the last step we have used (3.13) to replace $\bt_nR_{n-1}$ by $r_n^2/R_n.$ $\quad\quad\Box$

We note that $B > 0$ can always be satisfied for the proper range of $\lambda.$

The equation (3.9) states that $\al_n$ is linear in $R_n$ up to
a linear form in $n$, together with (3.10) and (3.16), when combined with say, (3.13)
provide us with a pair of non-linear difference equations satisfied by $r_n$
and $R_n.$ We state this in the next theorem.
\begin{theorem}
The quantities $r_n$ and $R_n$ satisfy the difference equations;
\bea
r_{n+1}+r_n&=&R_n(t-2n-\al-1-tR_n)\\
r_n^2\left(\frac{1}{R_nR_{n-1}}-\frac{1}{R_n}-\frac{1}{R_{n-1}}\right)
&=&r_n(2n+\al)+n(n+\al)
\eea
with the "initial" conditions,
\bea
r_0(t)&=&0\\
R_0(t)&=&\frac{B\:t^{\al}\re^{-t}}{h_0(t)}\\
h_0(t)&=&\left(1-\frac{\bt}{2}\right)^{\la}\Gamma(1+\al)+
\left[\left(1+\frac{\bt}{2}\right)^{\la}-\left(1-\frac{\bt}{2}\right)^{\la}\right]
\int_{t}^{\infty}x^{\al}\:\re^{-x}dx.
\eea
\end{theorem}

{\em Proof:} This is simply a restatement of (3.10) and (3.13) with (3.9) and (3.16)
$\quad\quad\Box$
\vskip .2cm

We shall see that $(S_2')$ automatically performs finite sums in "local" form, of the
quantities $R_n$ and $\al_n.$  This will be seen later to be relevant in the evaluation of
the derivative of $\ln D_n(t)$ with respect to $t$ and the derivation of the Painl\'eve
transcendent.

\begin{theorem}
\bea
t\sum_{j=0}^{n-1}R_j&=&-t\:r_n-n(n+\al)+\bt_n\\
\sum_{j=0}^{n-1}\al_j&=&-\textsf{p}_1(n)=\bt_n-t\:r_n.
\eea
\end{theorem}

{\em Proof:} From (3.15) we have,
\bea
t\sum_{j=0}^{n-1}R_j&=&r_n\left[2(n+r_n)+\al-t\right]+\bt_n\left[R_n+R_{n-1}-2R_nR_{n-1}\right]
\nonumber\\
&=&r_n\left[2(n+r_n)+\al-t\right]+\bt_n\left[R_n+R_{n-1}\right]-2r_n^2\nonumber\\
&=&r_n\left[2(n+r_n)+\al-t\right]+\bt_n-r_n(2n+\al)-n(n+\al)-2r_n^2\nonumber\\
&=&-t\:r_n-n(n+\al)+\bt_n. \eea The second equality of (3.26)
follows from (3.13) and the third equality follows from (3.17). The
equation (3.25) follows from (3.9) and the second equality of
(2.10). $\quad\quad\Box$

\setcounter{equation}{0}
\section{P$_V$(0,-$\frac{\al^2}{2}$,2n+1+$\al$,-$\frac{1}{2}$)}

In this section we shall discover which of the auxiliary quantities defined as the
residues of the rational functions $A_n(z)$ and $B_n(z)$ is a Painl\'eve transcendent.

This will be obtained from a pair of Toda equations which shows
that the Hankel determinant is the $\tau-$function and these when suitably combined
with the difference equations produce our $P_V.$

Taking the derivative of $h_n(t)$ with respect to $t,$ we find
\bea
\frac{d}{dt}\ln h_n(t)=-B\:w_0(t)\frac{\{P_n(t,t)\}^2}{h_n(t)}=-R_n(t),
\eea
and consequently we have the Theorem
\begin{theorem}
\bea
-t\frac{d}{dt}\ln D_n(t)&=&-t\sum_{j=0}^{n-1}\frac{d}{dt}\ln h_j(t)\nonumber\\
&=&t\sum_{j=0}^{n-1}R_j=-\textsf{p}_1(n,t)-n(n+\al).
\eea
\end{theorem}
{\em Proof:} The proof is obvious.$\quad\quad\Box$

The next lemma gives the derivative of $\textsf{p}_1(n,t)$ with respect to $t.$
\begin{lemma}
\bea
\frac{d}{dt}\textsf{p}_1(n,t)=r_n(t).
\eea
\end{lemma}

{\em Proof:} Note the $t$ dependence of $\textsf{p}_1(n,t).$
Taking a derivative of
$$
0=\int_{0}^{\infty}P_{n}(x)P_{n-1}(x)w_J(x,t)w_0(x)dx,
$$
with respect to $t,$ produces,
\bea
0&=&-B\:w_0(t)\:P_{n}(t,t)P_{n-1}(t, t)+
\int_{0}^{\infty}\left[\frac{d}{dt}\textsf{p}_1(n,t)x^{n-1}+...\right]P_{n-1}(x)
w_J(x,t)w_0(x)dx\nonumber\\
&=&-B\:w_0(t)P_{n}(t,t)P_{n-1}(t,t)+h_{n-1}\:\frac{d}{dt}\textsf{p}_1(n,t).\nonumber
\eea
and (4.3) follows.$\quad\quad\Box$

We expect $D_n(t)$ to satisfied the Toda molecule equation \cite{Sogo} and this should indicate the emergence
of a Painl\'eve transcendant. The question that we will address is "Which quantity is satisfied by
this particular Painl\'eve transcendant?"

\begin{theorem}
The Hankel determinant $D_n(t)$ satisfy the following differential-difference
or the Toda molecule equation \cite{Sogo},
\bea
t^2\frac{d^2}{dt^2}\ln D_n(t)=-n(n+\al)+\frac{D_{n+1}(t)D_{n-1}(t)}{D_n^2(t)}.
\eea
\end{theorem}

{\em Proof:} Taking a derivative of (4.2) with respect to $t$ and (4.3) imply
\bea
\frac{d}{dt}\left(t\frac{d}{dt}\ln D_n(t)\right)=r_n.\nonumber
\eea
Now substitute $r_n$ given above into (3.24) to find,
\bea
t\sum_{j=0}^{n-1}R_j&=&-t\frac{d}{dt}\left[t\frac{d}{dt}\ln D_n(t)\right]-n(n+\al)+\bt_n.\nonumber\\
&=&-t\frac{d}{dt}\ln D_n(t),\nonumber
\eea
where the last equality comes from (4.2). The equation (4.4) follows if we recall
$$
\bt_n=\frac{h_n}{h_{n-1}}=\frac{D_{n+1}D_{n-1}}{D_n^2},
$$
since $D_n=h_0...h_{n-1}.\quad\quad \Box$

We now state a pair of somewhat non-standard Toda equations.
\begin{lemma}
The recurrence coefficients $\al_n$ and $\bt_n$ satisfy for $n\in\{1,2,..\}$ the
differential-difference equations
$$
\bt_n'(t)=(R_{n-1}-R_n)\bt_n\eqno(T_1)
$$
$$
\al_n'(t)=r_n-r_{n-1},\eqno(T_2)
$$
with $r_0(t)$ and $R_0(t)$ given by (3.22) and (3.23) respectively.
\end{lemma}

{\em Proof:} These equations are an immediate consequence of (4.1), (2.11),
(4.3) and the first equality (2.10).$\quad\quad\Box$

To discover the $P_V$ of our problem. We first state two
preliminary lemmas describing the $t$ evolution of $r_n$ and $R_n.$

\begin{lemma}
For a fixed $n,$ $R_n(t)$ satisfies the Riccati equation,
\bea
tR_n'=2r_n+(2n+\al-t+tR_n)R_n.
\eea
\end{lemma}

{\em Proof:}
We begin with $(T_2)$ and replace $r_{n+1}$ by $R_n(t-\al_n)-r_n.$ See (3.7). This
leaves
\bea
\al_n'=2r_n-(t-\al_n)R_n\nonumber
\eea
After eliminating $\al_n$ in favor of $R_n$ with (3.9) we have (4.5). $\quad\quad\Box$

\begin{lemma}
For a fixed $n,\;\;r_n(t)$ satisfy the Riccati equation,
\bea
tr_n'=\frac{1-2R_n}{R_n(1-R_n)}\:(r_n)^2
-(2n+\al)\frac{R_nr_n}{1-R_n}-n(n+\al)\frac{R_n}{1-R_n}.
\eea
\end{lemma}

{\em Proof:} By equating (3.24) to the last equality of (4.2), we find
\bea
\textsf{p}_1(n,t)=tr_n-\bt_n.\nonumber
\eea
Taking a derivative of the above equation with respect to $t$ and noting (4.3)
we see that
\bea
tr_n'&=&\bt_n'\nonumber\\
&=&[R_{n-1}-R_n]\bt_n\nonumber\\
&=&\frac{r_n^2}{R_n}-\bt_n\:R_n,\nonumber
\eea
and use have been made of $(T_2)$ and (3.13) to obtain the last two equalities.
The equation (4.6) follows if we express $\bt_n$ in terms of $r_n$ and $R_n$ using (3.16).
$\quad\quad\Box$
\\
%
The next theorem shows that $R_n$ is up to a linear fractional transformation
a particular $P_V.$

\begin{theorem}
The quantity
\bea
S_n(t):=1-\frac{1}{R_n(t)},
\eea
satisfies
\beq{pain}
S_n''=\frac{3S_n-1}{2S_n(1-S_n)}\:(S_n')^2
-\frac{S_n'}{t}-\frac{\al^2}{2}\frac{(S_n-1)^2}{t^2\:S_n}
+(2n+1+\al)\frac{S_n}{t}-\frac{1}{2}\:\frac{S_n(S_n+1)}{S_n-1}.
\eeq
which is $P_V(0,-\al^2/2,2n+1+\al,-1/2).$

In terms of the recurrence coefficient $\al_n(t),$ we have,
\bea
S_n(t)=\frac{\al_n(t)-(2n+\al+1)-t}{\al_n(t)-(2n+\al+1)}.
\eea
\end{theorem}

{\em Proof:} Eliminate $r_n(t)$ from (4.5) and (4.6) and with $R_n=1/(1-S_n)$ gives
(4.8). We have followed the convention of \cite{Gromak}.$\quad\quad\Box$

\noindent
{\bf Remark 3.} Note that for $n=0,$ (4.8) is satisfied by
\bea
S_0(t)=1-\frac{1}{R_0(t)},\nonumber
\eea
where $R_0(t)$ is given by (3.22) and (3.23) and ultimately in terms of an Incomplete Gamma function---
a special case of the Kummer function of the second kind. Furthermore, since $r_0(t)=0,$ it can be
verified that $R_0(t)$ also satisfy (4.5) at $n=0.$

\vskip .3cm

We may express the logarithmic derivative of $D_n(t)$ with respect to $t$
in the so-called Jimbo-Miwa-Okamoto $\sigma$ form.
This is described in the next theorem.
\begin{theorem}
Let
\bea
H_n(t):=t\frac{d}{dt}\ln D_n(t),
\eea
then
\bea
(tH_n'')^2=4(H_n')^2[H_n-n(n+\al)-tH_n']+[(2n+\al-t)H_n'+H_n]^2.
\eea
\end{theorem}
{\em Proof:} First we express $r_n(t)$ and $\bt_n(t)$ in terms of $H_n$ and its derivatives.
From (3.24) and (4.2) we have
\bea
-H_n&=&-t\:r_n+\bt_n-n(n+\al)\nonumber\\
&=&-\textsf{p}_1(n,t)-n(n+\al).
\eea
Taking a derivative of (4.2) with respect to $t$ and recalling (4.3) we have
\bea
r_n=H_n',
\eea
and with the first equality of (4.12) and (4.13), we find,
\bea
\bt_n=tH_n'-H_n+n(n+\al).
\eea
Now a derivative of (4.14) with respect to $t$ and $(T_1)$ gives
\bea
(tH_n')'-H_n'&=&tH_n''\nonumber\\
&=&\bt_n'=(R_{n-1}-R_n)\bt_n\nonumber\\
&=&\frac{r_n^2}{R_n}-\bt_nR_n.
\eea
Here we have made use of (3.13) to arrive at the last equality.
Therefore we have a quadratic equation in $R_n;$
\bea
\frac{r_n^2}{R_n}-\bt_nR_n=tH_n''.
\eea
There is another quadratic equation in $R_n$ which is a restatement of (3.16);
\bea
\frac{r_n^2}{R_n}+\bt_nR_n=\bt_n-(2n+\al)r_n-n(n+\al).
\eea
Now we solve for $R_n$ and $1/R_n$ from (4.16) and (4.17) and find
\bea
\frac{2r_n^2}{R_n}&=&\bt_n-(2n+\al)r_n-n(n+\al)+tH_n''\nonumber\\
2\bt_nR_n&=&\bt_n-(2n+\al)r_n-n(n+\al)-tH_n''.\nonumber
\eea
The equation (4.10) follows from the product of the above two equations,
\bea
4\bt_nr_n^2=[\bt_n-(2n+\al)r_n-n(n+\al)]^2-(tH_n'')^2,\nonumber
\eea
and (4.13) and (4.14). $\quad\quad\Box$
\\

Incidentally $R_n$ has two alternative representations,
\bea
R_n&=&\frac{tH_n''+(2n+\al-t)H_n'+H_n}{2[H_n-n(n+\al)-tH_n']}\\
\frac{1}{R_n}&=&\frac{tH_n''-(2n+\al-t)H_n'-H_n}{2(H_n')^2}.
\eea
\\

The "discrete" structure inherited from the recurrence relations (2.7), induces a discrete analog
of the $\sigma$ form, namely, a non-linear second order difference equation in $n$ satisfied by
$H_n$ for a fixed $t$; we believe such a discrete form is new and
may have been missed in previous similar studies perhaps because the recurrence relations were not
sufficiently exploited. We note here that our derivation of (4.11) bypasses a third order
equation and without having to identify a first integral which reduces the order by one.

We note also that equation (4.11) was first discovered by Tracy and Widom in \cite{TW}
(which in our problem corresponds to $A=0$ and $B=1$) and just as was
done in their paper for the Hermite case one can also rescale to obtain the Painlev\'{e} III equation
corresponding to the Bessel kernel or ``hard edge scaling''. We  change variables $t \rightarrow s/4n,$
$H_{n} \rightarrow  \sigma$ use (4.11) and keep only the highest order terms to obtain
\[ (s\sigma^{\prime\prime})^{2} = 4\:\sigma(\sigma^{\prime})^{2}-4\:s(\sigma^{\prime})^{3} -s(\sigma^{\prime})
^{2}+\sigma\sigma^{\prime} +\alpha^{2}(\sigma^{\prime})^{2}.\]

Finally, we point out that the above analysis shows that the resolvent kernel used in the Tracy-Widom
approach can be directly related to the orthogonal polynomials defined on $(t, \infty)$. In fact,
if we denote $\tilde{R}(t,t)$ as the resolvent kernel defined in \cite{TW} then
\[t\tilde{R}(t,t) = H_{n}(t) = -tr_{n} -n(n+\alpha) + \beta_{n}.\]
Thus
\[t\tilde{R}(t,t) = -t\frac{Bw_{0}P_{n}(t,t)P_{n-1}(t,t)}{h_{n-1}(t)} -n(n+\alpha) + \frac{h_{n}(t)}{h_{n-1}(t)}.\]
The term $\beta_{n}$ can also be written using (3.16).
In addtion, we have that
\[\frac{d}{dt} \left(t\tilde{R}(t,t)\right) = r_{n} = \frac{Bw_{0}P_{n}(t,t)P_{n-1}(t,t)}{h_{n-1}(t)}.\]
In other words we have found an identity  for the resolvent kernel in terms of the values at the end points
of the normalized orthogonal polynomials.

\begin{theorem}
The auxiliary quantities  $R_n$ and $r_n$ are expressed in terms $H_n$ and $H_{n\pm 1}$
as follows:
\bea
t\:R_n&=&H_n-H_{n+1}\\
t\:r_n&=&\frac{[H_n-n(n+\al)](t+H_{n+1}-H_{n-1})+tn(n+\al)}{t+H_{n+1}-H_{n-1}-2n-\al}.
\eea
The discrete analog of the $\sigma$ form satisfied by $H_n$ results from the substitution of
(4.20) and (4.21) into
\bea
(t\:r_n)^2=[n(n+\al)+t\:r_n-H_n][(t\:R_n)^2+t\:R_n(H_{n+1}+H_{n-1}-2H_n)].
\eea
\end{theorem}

{\em Proof:} Taking a first order difference on the second equality of (4.2)
together with (2.10) and (3.9) implies (4.20).

We re-write (3.24) gives
\bea
\bt_n=n(n+\al)+t\:r_n-H_n.
\eea
We will now find another equation expressing $\bt_n$ in terms $r_n,\:R_n,\:H_n,\:$
$H_{n\pm 1}.$ Taking a first order difference on (4.20) gives,
\bea
t(R_n-R_{n-1})=2H_n-H_{n+1}-H_{n-1}.\nonumber
\eea
Now multiply the above equation by $R_n$ and make use of (3.13) we find
\bea
tR_n^2-\frac{tr_n^2}{\bt_n}=(2H_n-H_{n+1}-H_{n-1})R_n,\nonumber
\eea
and therefore
\bea
\frac{1}{\bt_n}=\frac{tR_n^2-(2H_n-H_{n+1}-H_{n-1})R_n}{t\:r_n^2}.
\eea
Therefore the product of (4.23) and (4.24) implies (4.21), which leaves us the job
of finding a further expression of $r_n$ in terms of $H_n\;$ and $H_{n\pm 1}.$
For this purpose we rewrite (3.17) as
\bea
\bt_n(1-R_n-R_{n-1})=(2n+\al)r_n+n(n+\al).\nonumber
\eea
Now substitute $\bt_n$ given in (4.23) into the above resulting a linear
equation in $r_n;$
\bea
r_n[(t-tR_{n}-tR_{n-1})-2n-\al]=[H_n-n(n+\al)](1-R_n-R_{n-1})+n(n+\al).\nonumber
\eea
With $t\:R_{n}$ as in (4.20) we have (4.21).

We summarize our results in the next theorem
\begin{theorem}
Let $D_n(t)$ be the Hankel determinant associated with the Laguerre weight perturbed by a jump factor,
and
$$
H_n(t):=t\frac{d}{dt}\ln D_n(t).
$$
Then the recurrence coefficients are
\bea
\al_n(t)-(2n+\al+1)&=&\frac{t^2H_n''+[(2n+\al)t-t^2]H_n'+tH_{n}}{2[H_n-n(n+\al)-tH_n']}\\
\bt_n(t)-n(n+\al)&=&tH_n'-H_n,
\eea
where $2n+1+\al$ and $n(n+\al)$ are the "unperturbed" recurrence coefficients
and $H_n$ satisfies a non-linear differential equation in Jimbo-Miwa-Okamoto $\sigma$ form,
\bea
(tH_n'')^2=4(H_n')^2[H_n-n(n+\al)-tH_n']+[(2n+\al-t)H_n'+H_n]^2.\nonumber
\eea
\noindent
 For the same $H_n,$ the recurrence coefficients are
 \bea
 \al_n(t)-(2n+\al+1)&=&H_n-H_{n+1}\\
\bt_n(t)-n(n+\al)&=&
\frac{H_n(2n+\al)-n(n+\al)(H_{n+1}-H_{n-1})}{t+H_{n+1}-H_{n-1}-2n-\al},
\eea
where $H_n$ satisfies the discrete $\sigma$ form of a non-linear difference equation,
\bea
&&\Bigg\{\frac{[H_n-n(n+\al)](t+H_{n+1}-H_{n-1})+t\:n(n+\al)}{t+H_{n+1}-H_{n-1}-2n-\al}\Bigg\}^2
\nonumber\\
&=&
\Bigg\{\frac{(2n+\al)[H_n-n(n+\al)]+tn(n+\al)}{t+H_{n+1}-H_{n-1}-2n-\al}\Bigg\}
(H_{n}-H_{n+1})(H_{n-1}-H_n).
\eea
\end{theorem}
Note that since$\al_n(t)$ and
$\bt_n(t)$ have two alternative representations,$H_n(t)$ satisfies two more differential-difference
equations, $(4.25)=(4.27)$ and $(4.26)=(4.28).$
\vskip .2cm
We end this paper with a discussion on the relationship between our $P_V$  and the
difference equations (3.19) and (3.20). We would like to
thank the second referee for supplying us the background material part of which is reproduced here.
\vskip .2cm
The fifth Painlev\'e equation $P_V(a,b,c,d=-1/2):$
$$
y''=\left(\frac{1}{2y}+\frac{1}{y-1}\right)(y')^2-\frac{y'}{t}+
\frac{(y-1)^2}{t^2}\left(ay+\frac{b}{y}\right)+c\frac{y}{t}+d\frac{y(y+1)}{y-1},\:\:'=\frac{d}{dt}
$$
is equivalent to the Hamiltonian system ${\cal H}_V:$
$$
q'=\frac{\partial\H}{\partial p},\quad tp'=-\frac{\partial\H}{\partial q},
$$
with the time-dependent Hamiltonian $\H=\H(p,q,t):$
$$
t\H=p(p+t)q(q-1)+\al_2qt-\al_3pq-\al_1p(q-1),
$$
where
$$
a=\frac{\al_1^2}{2},\;\;b=-\frac{\al_3^2}{2},\;\;c=\al_0-\al_2,\;\;\;d=-\frac{1}{2},\;\;
\al_0:=1-\al_1-\al_2-\al_3
$$
and
$$
y=1-\frac{1}{q}.
$$
The Hamiltonian structure was studied in \cite{Oka} and the $\tau-$function is defined such that
$$
\frac{d}{dt}\ln\tau=\H.
$$
The extended affine Weyl group $W(A_3^{(1)})=<s_0,s_1,s_2,s_3,\pi>$
of the Weyl group type $A_3^{(1)}$ acts
as bi-rational symmetries on $P_V$ and induces Backlund transformations on
the solutions of $P_V.$ Here the $s_i'$s and $\pi$ are the generators.
See \cite{Oka} for the study of Weyl group actions on $P_V$.

For example, the action of $s_0:$
\bea
s_0\{\al_0,\al_1,\al_2,\al_3\}&=&\{-\al_0,\al_1+\al_0,\al_2,\al_3+\al_0\},\nonumber\\
s_0(q)&=&q+\frac{\al_0}{p+t},\nonumber\\
s_0(p)&=&p,\nonumber
\eea
leaves $P_V$ or the Hamiltonian system ${\cal H}_V$ invariant. We refer the readers to
\cite{Oka,Tsuda} for information on Weyl group actions and [(4.3),\cite{Tsuda}] which lists the
bi-rational transformations.

To proceed further, consider a parallel transformation $l=(s_2s_3\pi)^2\in W(A_3^{(1)}):$
\bea
l:\vec{\al}=(\al_0,\al_1,\al_2,\al_3)\longmapsto\vec{\al}+(1,0,-1,0).\nonumber
\eea
From a direct computation, we may verify that, the variables $q$ and $r:=pq(q-1)$
satisfies the following system of difference equations:
\bea
l(r)+r&=&q(\al_2-\al_0+t-tq)-\al_1\\
\left(\frac{1}{q}-1\right)\left(\frac{1}{l^{-1}(q)}-1\right)&=&\frac{(r-\al_2)(r-\al_2-\al_3)}{r(r+\al_1)},
\eea
these seems to the $d-P_{\rm III}$ of [205,\cite{Sakai}] in disguise.

In terms of $\H$ our auxiliary parameter $r$ reads
\bea
r=\frac{d}{dt}\left(t\H\right).
\eea

In our problem we have $P_V(0,-\al^2/2,2n+1+\al,-1/2),$ which implies
$$
(\al_0,\al_1,\al_2,\al_3)=(1+n,0,-n-\al,\al).
$$
If
\bea
l^{n}(q)&=&q_n=:R_n\nonumber\\
l^{n}(r)&=:&r_n,\nonumber
\eea
for $n\in\{0,1,2,...\},$ then a direct computation shows that (4.33) and (4.34) are
\bea
r_{n+1}+r_n&=&R_n(-\al-2n-1+t-tR_n),\\
\left(\frac{1}{R_n}-1\right)\left(\frac{1}{R_{n-1}}-1\right)&=&\frac{(r_n+n+\al)(r_n+n)}{r_n^2},
\eea
are equivalent to (3.19) and (3.20) respectively.

We should like to mention here that (3.19) and (3.20) and other equations are derived
entirely from orthogonality and the immediate consequence---the recurrence relations.

 In view of
(3.32) we see that the logarithmic derivative of the generating function $\textsf{G}(n,t)=D_n(t)$
is the $\tau-$ function of our $P_V.$ We end this paper with the final remark: the equation
(4.4) is essentially the same as the Toda equation among a $\tau-$sequence discovered by
Okamoto \cite{Oka}.


\end{document}